\documentclass[reprint,runinaddress,frontmatterverbose,aps,showkeys, showpacs,prb,floatfix,superscriptaddress]{revtex4-2}
\usepackage{graphicx}
\usepackage{dcolumn}
\usepackage{bm}
\usepackage{graphicx}
\usepackage{color}
\usepackage{colortbl}
\usepackage[dvipsnames]{xcolor} 
\usepackage{tabularx}
\usepackage{epsfig}
\usepackage{amsmath}
\usepackage{graphicx}
\usepackage{dcolumn,epstopdf}
\usepackage{bm}
\usepackage{wasysym}
\usepackage{times}
\usepackage{float}
\usepackage{gensymb}
\definecolor{grn}{rgb}{0,0,0.54}

\usepackage{latexsym}
\usepackage{amsfonts}
\usepackage{amsmath}
\def\be{\begin{equation}}
\def\ee{\end{equation}}
\newcommand{\bea}{\begin{eqnarray}}
\newcommand{\eea}{\end{eqnarray}}

\begin{document}


\title[Tunnel diode oscillator]{A Tunnel Diode Oscillator for High Sensitivity Broad Temperature Range Susceptibility Measurements}

\author{A. Sirusi}
\author{J. Adams}%
 
\author{M. Lewkowitz}
\author{ R. Sun}
\author{N. S. Sullivan}
\email{Sullivan@phys.ufl.edu}
\affiliation{ 
Department of Physics, University of Florida, Gainesville, FL32611, USA.
}%


\date{\today}

\begin{abstract}
We report the design and operation of a  versatile tunnel diode oscillator for high sensitivity ( $\approx
50$ ppb) measurements of the magnetic susceptibility of samples over a wide temperature
range   (1.7-100 K) that can be used with a simple liquid helium storage dewar. The design allows for the application of an electric field across the sample
to search for magnetoelectric effects.
\end{abstract}

\maketitle

\section{\label{sec:level1}Introduction}
We needed a versatile and robust system for measuring magnetic susceptibilities over a wide temperature range ({\color{black} 1.7 - 100 K) with high sensitivity.} A number of tunnel diode oscillators have been built for limited purposes, either with high sensitivity (better than 100 ppm) over a limited range of temperatures,
 \cite{Degrift1981,Drigo2010} 
 or with  low sensitivity ($\geq 1000$ ppm) but capable of functioning over a broad range of external parameters (temperature and pressure).
\cite{Clover1970,Srikanth1999,Agosta2000,Mielke2009,Komatsu2004}

The design reported here 
 maintains the tunnel diode at a fixed low temperature (2 - 4K) while the sample is located in a separate resonant coil of Helmholtz design that can be varied in temperature {\color{black} up to}  100 K. The Helmholtz design allows the placement of electrodes across the sample (typically mm in dimension) providing static electric fields (up to $10^5$ V/m). AC electric fields applied at low frequency
can be used to detect magnetoelectric effects synchronously with the electric excitation. This feature is important as it has been shown that the magnetoelectric effects can have a strong frequency dependence.
\cite{Xia2017}

\section{Design Considerations}

Figure 1 provides a schematic representation of the system. The tunnel diode and its biasing circuitry is located on a small copper clad circuit board (\color{black}TD \color{black} block in Fig.\,1) at the lowest end of the interior cryostat and maintained at a fixed low temperature
{\color{black} using a hard pressed thermal contact via  a  spring loaded clip to  bath B containing} superfluid helium.  A compression force of the order of 100 N  can provide a cooling power of a few tens of mW at helium temperatures.\cite{GKWhite} The superfuid helium bath B is supplied by a  flow limiting impedance (Z $\approx  10^{10}$ cm$^{-3}$) that is connected to a normal (4.2 K) helium bath, and has a maximum cooling capacity at T $\approx 2$ K of $\sim$ 100 mW. The volume flow ${\dot V}_4 = \Delta P /(\eta Z)$ where $\eta$ is the viscosity of the helium (in Poise) and $\Delta P \approx ~ 10^5 $ Pa is the pressure drop across Z.\cite{DeLong1971} The filter F screens out small particles of frozen air that might be suspended in the liquid helium and could block the impedance Z. {\color{black} The entire assembly can be inserted in simple wide neck liquid helium storage dewar and thereby keeping the helium consumption to a low value.} 


\begin{figure}[ht!]
\includegraphics[width=12pc]{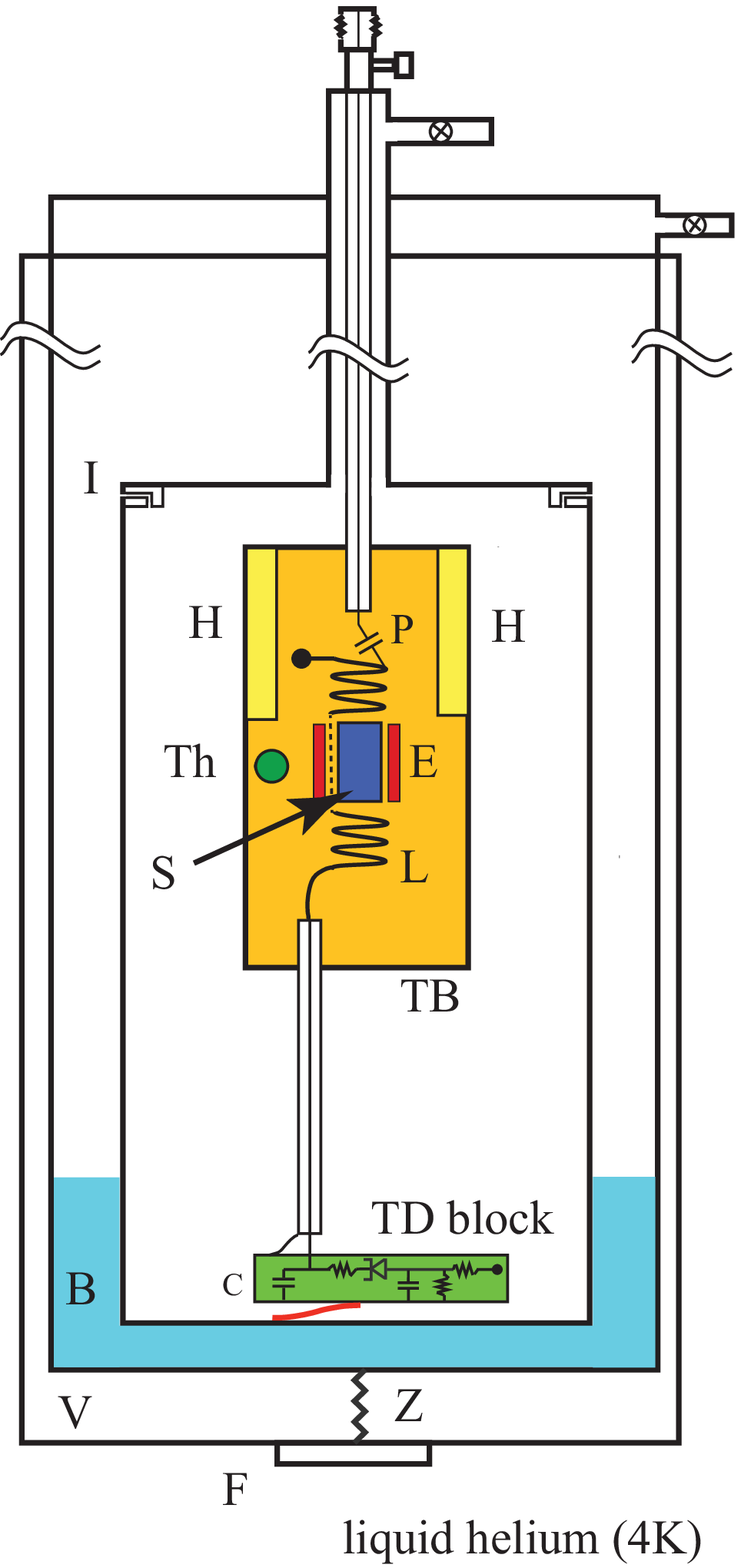} 
\caption{\label{schematic}
Schematic representation of the low temperature apparatus (not to scale). The tunnel diode is held at constant temperature by a spring contact  (red) at the bottom of the inner container which is the wall of the pumped liquid helium bath B. The sample S is located between Helmholtz windings L which allow for two electrodes E that apply an electric field across the sample. The thermal board TB is maintained at a fixed temperature by means of a feedback heater, the thermal connection to the  TB block being very weak.  The outer dimension was 2.25 in and the entire assembly could be inserted into a simple wide neck liquid helium storage dewar for ease of operation. 
}
\end{figure}

Maintaining the tunnel diode (General Electric type BD4)\cite{Lowry} at a fixed temperature maximizes the stability of the frequency and amplitude of the oscillations and allows one to carefully select the operating point for which the frequency variations are a minimum with respect to bias voltage as {\color{black} noted} by Clover and Wolf\cite{Clover1970}). The {\color{black}   sample cell was located at the center of a quasi-Helmholtz radiofrequency coil L that was thermally anchored} to an independent thermal board whose temperature could be carefully regulated by a local heater. The sample S was contained in a small {\color{black} Kel-F} container and thermally bonded to the thermal board with N-grease.{\color{black}\cite{Apiezon}} The inductance L was resonated with a silver-dipped mica capacitor C located on the TD block and, depending on choice of the coil L could be operated between 2 and 120 MHz. \color{black}The sample cell block was shielded by a copper cylinder that minimizes external noise and provides {\color{black} additional} thermal isolation for the sample. \color{black}


The fixed temperature tunnel diode block is separated from the sample cell block by a 10 cm length of semi-rigid cryogenic coaxial cable with stainless-steel outer conductor and silver plated beryllium-copper inner conductor\cite{Coaxitube} that provides adequate thermal isolation  to maintain the sample board at a temperature between {\color{black} 1.7 K and 100 K} (with a precision of better than 1 \%) by using a  regulated feedback circuit\color{black}, while the tunnel diode block is held at liquid helium temperatures \color{black}. The \color{black}sample cell and tunnel diode block \color{black} temperatures were measured {\color{black} using calibrated}  Cernox thermometers.{\color{black}\cite{Lakeshore}} A small amount of $^4$He exchange gas was added to the sample space to provide additional cooling.

  As shown in figure 2, the {\color{black} open} Helmholtz coil design allowed one to introduce small electrodes to place \color{black} a static \color{black} electric field (up to 100 kV.m$^{-1}$) across the sample to detect magnetoelectric effects. In addition to a static field a small AC field can be applied and used in conjuction with FM detection for high resolution, ($\sim$ hundreds of ppb), detection of magnetoelectric effects.
	
\begin{figure}[ht]
\includegraphics[width=10pc]{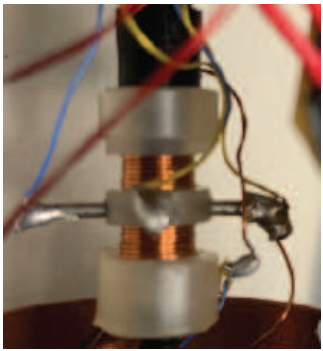}
\caption{\label{Cell} Picture of {\color{black} Kel-F} sample cell showing the Helmholtz coil pair (5 mm diameter) and the electrodes for generating an electric field across the sample (typically 1 mm in lateral dimension. The sample is thermalized by applying Apiezon N-grease which makes contact with the thermal board.
}
\end{figure}


\section{Performance}

{\color{black} In order to test the sensitivity we used a commercial sample of MgB${_2}$ \cite{Sigma} that had a known magnetic susceptibility change at the superconducting transition temperature of 40.05$\pm$ 0.10 K (in good agreement with the values for the critical temperature reported in the literature.\cite{Buzea_2001} The fractional  change in oscillator frequency
\begin{equation}
    \frac{\Delta f}{f} =-\frac{1}{2}\eta \chi^{'} +O(Q^{-2}\chi^{"})
    \end{equation}
    where $\eta$ is the sample filling factor and $\chi^{'}$ and $\chi^{"}$ are the real and imaginary components of the sample susceptibility, respectively.
is in good agreement with the estimated value of {\color{black} $\eta \sim 4$\%.
 The noise level in the low resolution channel was  12ppm and in the high resolution AC channel was of the order of 0.01 ppm, comparable to the best performance reported by Degrift\cite{Degrift1975} for a low current diode BD6 at a fixed temperature.} Similar sensitivity was reported by Van Riet and Van Gerver who used a low temperature tunnel diode circuit to observe NMR in a micromolar sample at 4K.\cite{Riet1982}
 
\begin{figure}[h]
\hbox{\hspace{+0.6em}\includegraphics[width=25pc]{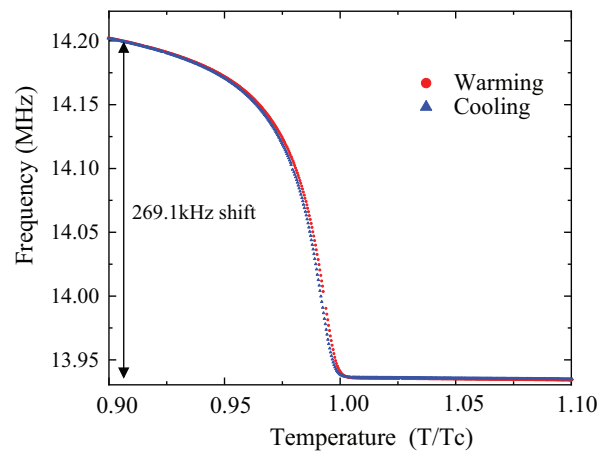}}
\caption{\label{MgB2} Frequency response {\color{black} measured for the magnetic susceptibility} of high-temperature superconductor MgB$_2$.{\color{black}  The fractional change in frequency corresponds to a filling factor of 3.8\%.} } 
\end{figure}



\section{\label{sec:level3}Conclusion}
We have demonstrated the use of a novel cryogenic design for a tunnel diode oscillator for measuring the magnetic susceptibility of small samples over a wide temperature range (1.7-100 K). High sensitivity and high stability were achieved by thermally separating the sample cell from the tunnel diode with the latter held at fixed helium temperatures. The compact geometry allows on to insert the complete system in a simple liquid helium dewar keeping the boil off to less than 0.2 liquid liters per hour at the highest sample temperatures.

\begin{acknowledgments}
We would like to thank the University of Florida Cryogenics Facility and the Machine and Electronic shops of the Physics Department for their assistance with this project. This work was supported as part of the Center for Molecular Magnetic Quantum Materials, an Energy Frontier Research Center funded by the U.S. Department of Energy, Office of Science, Basic Energy Sciences under Award \# DE-SC0019330.

\end{acknowledgments}
\bibliography{TDO_bib.tex}

\begin{thebibliography}{18}%
\makeatletter
\providecommand \@ifxundefined [1]{%
 \@ifx{#1\undefined}
}%
\providecommand \@ifnum [1]{%
 \ifnum #1\expandafter \@firstoftwo
 \else \expandafter \@secondoftwo
 \fi
}%
\providecommand \@ifx [1]{%
 \ifx #1\expandafter \@firstoftwo
 \else \expandafter \@secondoftwo
 \fi
}%
\providecommand \natexlab [1]{#1}%
\providecommand \enquote  [1]{``#1''}%
\providecommand \bibnamefont  [1]{#1}%
\providecommand \bibfnamefont [1]{#1}%
\providecommand \citenamefont [1]{#1}%
\providecommand \href@noop [0]{\@secondoftwo}%
\providecommand \href [0]{\begingroup \@sanitize@url \@href}%
\providecommand \@href[1]{\@@startlink{#1}\@@href}%
\providecommand \@@href[1]{\endgroup#1\@@endlink}%
\providecommand \@sanitize@url [0]{\catcode `\\12\catcode `\$12\catcode
  `\&12\catcode `\#12\catcode `\^12\catcode `\_12\catcode `\%12\relax}%
\providecommand \@@startlink[1]{}%
\providecommand \@@endlink[0]{}%
\providecommand \url  [0]{\begingroup\@sanitize@url \@url }%
\providecommand \@url [1]{\endgroup\@href {#1}{\urlprefix }}%
\providecommand \urlprefix  [0]{URL }%
\providecommand \Eprint [0]{\href }%
\providecommand \doibase [0]{https://doi.org/}%
\providecommand \selectlanguage [0]{\@gobble}%
\providecommand \bibinfo  [0]{\@secondoftwo}%
\providecommand \bibfield  [0]{\@secondoftwo}%
\providecommand \translation [1]{[#1]}%
\providecommand \BibitemOpen [0]{}%
\providecommand \bibitemStop [0]{}%
\providecommand \bibitemNoStop [0]{.\EOS\space}%
\providecommand \EOS [0]{\spacefactor3000\relax}%
\providecommand \BibitemShut  [1]{\csname bibitem#1\endcsname}%
\let\auto@bib@innerbib\@empty
\bibitem [{\citenamefont {Van~Degrift}\ and\ \citenamefont
  {Love}(1981)}]{Degrift1981}%
  \BibitemOpen
  \bibfield  {author} {\bibinfo {author} {\bibfnamefont {C.~T.}\ \bibnamefont
  {Van~Degrift}}\ and\ \bibinfo {author} {\bibfnamefont {D.~P.}\ \bibnamefont
  {Love}},\ }\href {https://doi.org/10.1063/1.1136656} {\bibfield  {journal}
  {\bibinfo  {journal} {Review of Scientific Instruments}\ }\textbf {\bibinfo
  {volume} {52}},\ \bibinfo {pages} {712} (\bibinfo {year} {1981})},\ \Eprint
  {https://arxiv.org/abs/https://doi.org/10.1063/1.1136656}
  {https://doi.org/10.1063/1.1136656} \BibitemShut {NoStop}%
\bibitem [{\citenamefont {Drigo}\ \emph {et~al.}(2010)\citenamefont {Drigo},
  \citenamefont {Durantel}, \citenamefont {Audouard},\ and\ \citenamefont
  {Ballon}}]{Drigo2010}%
  \BibitemOpen
  \bibfield  {author} {\bibinfo {author} {\bibfnamefont {L.}~\bibnamefont
  {Drigo}}, \bibinfo {author} {\bibfnamefont {F.}~\bibnamefont {Durantel}},
  \bibinfo {author} {\bibfnamefont {A.}~\bibnamefont {Audouard}},\ and\
  \bibinfo {author} {\bibfnamefont {G.}~\bibnamefont {Ballon}},\ }\href
  {https://doi.org/10.1051/epjap/2010127} {\bibfield  {journal} {\bibinfo
  {journal} {Eur. Phys. J. Appl. Phys.}\ }\textbf {\bibinfo {volume} {52}},\
  \bibinfo {pages} {10401} (\bibinfo {year} {2010})}\BibitemShut {NoStop}%
\bibitem [{\citenamefont {Clover}\ and\ \citenamefont
  {Wolf}(1970)}]{Clover1970}%
  \BibitemOpen
  \bibfield  {author} {\bibinfo {author} {\bibfnamefont {R.~B.}\ \bibnamefont
  {Clover}}\ and\ \bibinfo {author} {\bibfnamefont {W.~P.}\ \bibnamefont
  {Wolf}},\ }\href {https://doi.org/10.1063/1.1684598} {\bibfield  {journal}
  {\bibinfo  {journal} {Rev. Sci. Instr.}\ }\textbf {\bibinfo {volume} {41}},\
  \bibinfo {pages} {617} (\bibinfo {year} {1970})}\BibitemShut {NoStop}%
\bibitem [{\citenamefont {Srikanth}\ \emph {et~al.}(1999)\citenamefont
  {Srikanth}, \citenamefont {Wiggins},\ and\ \citenamefont
  {Rees}}]{Srikanth1999}%
  \BibitemOpen
  \bibfield  {author} {\bibinfo {author} {\bibfnamefont {H.}~\bibnamefont
  {Srikanth}}, \bibinfo {author} {\bibfnamefont {J.}~\bibnamefont {Wiggins}},\
  and\ \bibinfo {author} {\bibfnamefont {H.}~\bibnamefont {Rees}},\ }\href
  {https://doi.org/10.1063/1.1149892} {\bibfield  {journal} {\bibinfo
  {journal} {Rev. . Sci. Instr.}\ }\textbf {\bibinfo {volume} {70}},\ \bibinfo
  {pages} {3097} (\bibinfo {year} {1999})}\BibitemShut {NoStop}%
\bibitem [{\citenamefont {Coffey}\ \emph {et~al.}(2000)\citenamefont {Coffey},
  \citenamefont {Bayindir}, \citenamefont {DeCarolis}, \citenamefont {Bennett},
  \citenamefont {Esper},\ and\ \citenamefont {Agosta}}]{Agosta2000}%
  \BibitemOpen
  \bibfield  {author} {\bibinfo {author} {\bibfnamefont {T.}~\bibnamefont
  {Coffey}}, \bibinfo {author} {\bibfnamefont {Z.}~\bibnamefont {Bayindir}},
  \bibinfo {author} {\bibfnamefont {J.~F.}\ \bibnamefont {DeCarolis}}, \bibinfo
  {author} {\bibfnamefont {M.}~\bibnamefont {Bennett}}, \bibinfo {author}
  {\bibfnamefont {G.}~\bibnamefont {Esper}},\ and\ \bibinfo {author}
  {\bibfnamefont {C.~C.}\ \bibnamefont {Agosta}},\ }\href
  {https://doi.org/10.1063/1.1321301} {\bibfield  {journal} {\bibinfo
  {journal} {Rev.Sci. Instr.}\ }\textbf {\bibinfo {volume} {71}},\ \bibinfo
  {pages} {4600} (\bibinfo {year} {2000})}\BibitemShut {NoStop}%
\bibitem [{\citenamefont {Altarawneh}\ \emph {et~al.}(2009)\citenamefont
  {Altarawneh}, \citenamefont {Mielke},\ and\ \citenamefont
  {Brooks}}]{Mielke2009}%
  \BibitemOpen
  \bibfield  {author} {\bibinfo {author} {\bibfnamefont {M.~M.}\ \bibnamefont
  {Altarawneh}}, \bibinfo {author} {\bibfnamefont {C.~H.}\ \bibnamefont
  {Mielke}},\ and\ \bibinfo {author} {\bibfnamefont {J.~S.}\ \bibnamefont
  {Brooks}},\ }\href {https://doi.org/10.1063/1.3152219} {\bibfield  {journal}
  {\bibinfo  {journal} {Rev..Sci. Instr.}\ }\textbf {\bibinfo {volume} {80}},\
  \bibinfo {pages} {066104} (\bibinfo {year} {2009})}\BibitemShut {NoStop}%
\bibitem [{\citenamefont {Komatsu}\ \emph {et~al.}(2004)\citenamefont
  {Komatsu}, \citenamefont {Ohmichi},\ and\ \citenamefont
  {Osada}}]{Komatsu2004}%
  \BibitemOpen
  \bibfield  {author} {\bibinfo {author} {\bibfnamefont {E.}~\bibnamefont
  {Komatsu}}, \bibinfo {author} {\bibfnamefont {E.}~\bibnamefont {Ohmichi}},\
  and\ \bibinfo {author} {\bibfnamefont {T.}~\bibnamefont {Osada}},\ }\href
  {https://doi.org/https://doi.org/10.1016/j.physb.2004.01.142} {\bibfield
  {journal} {\bibinfo  {journal} {Physica B: Condensed Matter}\ }\textbf
  {\bibinfo {volume} {346-347}},\ \bibinfo {pages} {534 } (\bibinfo {year}
  {2004})},\ \bibinfo {note} {{P}roceedings of the 7th International Symposium
  on Research in High Magnetic Fields}\BibitemShut {NoStop}%
\bibitem [{\citenamefont {Xia}\ \emph {et~al.}(2017)\citenamefont {Xia},
  \citenamefont {Yin}, \citenamefont {Sullivan}, \citenamefont {Zapf},\ and\
  \citenamefont {Paduan-Filho}}]{Xia2017}%
  \BibitemOpen
  \bibfield  {author} {\bibinfo {author} {\bibfnamefont {J.~S.}\ \bibnamefont
  {Xia}}, \bibinfo {author} {\bibfnamefont {L.}~\bibnamefont {Yin}}, \bibinfo
  {author} {\bibfnamefont {N.~S.}\ \bibnamefont {Sullivan}}, \bibinfo {author}
  {\bibfnamefont {V.~S.}\ \bibnamefont {Zapf}},\ and\ \bibinfo {author}
  {\bibfnamefont {A.}~\bibnamefont {Paduan-Filho}},\ }\href
  {https://doi.org/10.1007/s10909-016-1723-5} {\bibfield  {journal} {\bibinfo
  {journal} {J. Low Temp. Phys.}\ }\textbf {\bibinfo {volume} {187}},\ \bibinfo
  {pages} {627} (\bibinfo {year} {2017})}\BibitemShut {NoStop}%
\bibitem [{\citenamefont {White}\ and\ \citenamefont {Meeson}(2002)}]{GKWhite}%
  \BibitemOpen
  \bibfield  {author} {\bibinfo {author} {\bibfnamefont {G.~K.}\ \bibnamefont
  {White}}\ and\ \bibinfo {author} {\bibfnamefont {P.~J.}\ \bibnamefont
  {Meeson}},\ }\href@noop {} {\emph {\bibinfo {title} {Experimental Techniques
  in Low Temeprature Physics}}}\ (\bibinfo  {publisher} {Clarendon Press},\
  \bibinfo {address} {Oxford University, {UK}},\ \bibinfo {year}
  {2002})\BibitemShut {NoStop}%
\bibitem [{\citenamefont {DeLong}\ \emph {et~al.}(1971)\citenamefont {DeLong},
  \citenamefont {Symko},\ and\ \citenamefont {Wheatley}}]{DeLong1971}%
  \BibitemOpen
  \bibfield  {author} {\bibinfo {author} {\bibfnamefont {L.~E.}\ \bibnamefont
  {DeLong}}, \bibinfo {author} {\bibfnamefont {O.~G.}\ \bibnamefont {Symko}},\
  and\ \bibinfo {author} {\bibfnamefont {J.~C.}\ \bibnamefont {Wheatley}},\
  }\href {https://doi.org/10.1063/1.1684846} {\bibfield  {journal} {\bibinfo
  {journal} {Rev. Sci. Instr.}\ }\textbf {\bibinfo {volume} {42}},\ \bibinfo
  {pages} {147} (\bibinfo {year} {1971})}\BibitemShut {NoStop}%
\bibitem [{\citenamefont {Lowry}\ \emph {et~al.}(1961)\citenamefont {Lowry},
  \citenamefont {Giorgis}, \citenamefont {Gotlieb},\ and\ \citenamefont
  {Weischedel}}]{Lowry}%
  \BibitemOpen
  \bibfield  {author} {\bibinfo {author} {\bibfnamefont {H.~R.}\ \bibnamefont
  {Lowry}}, \bibinfo {author} {\bibfnamefont {J.}~\bibnamefont {Giorgis}},
  \bibinfo {author} {\bibfnamefont {E.}~\bibnamefont {Gotlieb}},\ and\ \bibinfo
  {author} {\bibfnamefont {R.~C.}\ \bibnamefont {Weischedel}},\ }\href@noop {}
  {\bibinfo {title} {Type 4jf2a-4, {T}unnel {D}iode {M}anual ({G}eneral
  {E}lectric)}} (\bibinfo {year} {1961})\BibitemShut {NoStop}%
\bibitem [{Api()}]{Apiezon}%
  \BibitemOpen
  \href@noop {} {\bibinfo {title} {Apiezon type {N}-grease}},\ \bibinfo {note}
  {{M}\& {I} {M}aterials {L}td, Manchester, UK}\BibitemShut {NoStop}%
\bibitem [{Coa()}]{Coaxitube}%
  \BibitemOpen
  \href@noop {} {\bibinfo {title} {{C}oaxitube {T}ype {JN}50141}},\ \bibinfo
  {note} {precision {T}ube Co.}\BibitemShut {Stop}%
\bibitem [{Lak()}]{Lakeshore}%
  \BibitemOpen
  \href@noop {} {\bibinfo {title} {Cernox type 1030}},\ \bibinfo {note}
  {lakeshore {C}ryotronics}\BibitemShut {NoStop}%
\bibitem [{Sig()}]{Sigma}%
  \BibitemOpen
  \href@noop {} {\bibinfo {title} {{P}roduct 553913}},\ \bibinfo {note}
  {{S}igma {A}ldrich}\BibitemShut {NoStop}%
\bibitem [{\citenamefont {Buzea}\ and\ \citenamefont
  {Yamashita}(2001)}]{Buzea_2001}%
  \BibitemOpen
  \bibfield  {author} {\bibinfo {author} {\bibfnamefont {C.}~\bibnamefont
  {Buzea}}\ and\ \bibinfo {author} {\bibfnamefont {T.}~\bibnamefont
  {Yamashita}},\ }\href {https://doi.org/10.1088/0953-2048/14/11/201}
  {\bibfield  {journal} {\bibinfo  {journal} {Supercon. Sci. and Tech.}\
  }\textbf {\bibinfo {volume} {14}},\ \bibinfo {pages} {R115} (\bibinfo {year}
  {2001})}\BibitemShut {NoStop}%
\bibitem [{\citenamefont {Van~Degrift}(1975)}]{Degrift1975}%
  \BibitemOpen
  \bibfield  {author} {\bibinfo {author} {\bibfnamefont {C.~T.}\ \bibnamefont
  {Van~Degrift}},\ }\href {https://doi.org/10.1063/1.1134272} {\bibfield
  {journal} {\bibinfo  {journal} {Rev. Sci. Instr.}\ }\textbf {\bibinfo
  {volume} {46}},\ \bibinfo {pages} {599} (\bibinfo {year} {1975})}\BibitemShut
  {NoStop}%
\bibitem [{\citenamefont {Riet}\ and\ \citenamefont {Gerven}(1982)}]{Riet1982}%
  \BibitemOpen
  \bibfield  {author} {\bibinfo {author} {\bibfnamefont {B.~V.}\ \bibnamefont
  {Riet}}\ and\ \bibinfo {author} {\bibfnamefont {L.~V.}\ \bibnamefont
  {Gerven}},\ }\href {https://doi.org/10.1088/0022-3735/15/5/019} {\bibfield
  {journal} {\bibinfo  {journal} {Journal of Physics E: Scientific
  Instruments}\ }\textbf {\bibinfo {volume} {15}},\ \bibinfo {pages} {558}
  (\bibinfo {year} {1982})}\BibitemShut {NoStop}%
\end{thebibliography}%
\end{document}